\documentclass[10pt, conference]{IEEEtran}
\usepackage{cite, graphicx, subfigure,url,setspace,algorithm}
\usepackage{amsmath,amssymb}
\usepackage{bm}
\usepackage{subfigure}
\usepackage{CJK}

\newtheorem{remark}{Remark}

\usepackage{epstopdf}

\begin{document}
\title{ MULTI-OBJECT TRACKING WITH MULTIPLE BIRTH, DEATH, AND SPAWN SCENARIOS USING A RANDOMIZED HYPOTHESIS GENERATION TECHNIQUE (R-FISST)}

\author{
\IEEEauthorblockN{W. Faber and S. Chakravorty}
\IEEEauthorblockA{Department of Aerospace Engineering \\ Texas
  A\&M University \\ College Station, TX} 
  \and
\IEEEauthorblockN{Islam I. Hussein}\IEEEauthorblockA{Applied Defense Solutions \\ Columbia, MD}}
\maketitle

\setlength{\parskip}{0ex}

\begin{abstract}
In multi-object tracking one may encounter situations were at any time step the number of possible hypotheses is too large to generate exhaustively. These situations generally occur when there are multiple ambiguous measurement returns that can be associated to many objects. This paper contains a newly developed approach that keeps the aforementioned situations computationally tractable. Utilizing a hypothesis level derivation of the Finite Set Statistics (FISST) Bayesian recursions for multi-object tracking we are able to propose a randomized method called randomized FISST (R-FISST). Like our previous methods \cite{Faber1, Faber2}, this approach utilizes Markov Chain Monte Carlo (MCMC) methods to sample highly probable hypotheses, however, the newly developed (R-FISST) can account for hypotheses containing multiple births and death within the MCMC sampling.  This alleviates the burden of having to exhaustively enumerate all birth and death hypotheses and makes the method more equipped to handle spawn scenarios. We test our method on Space Situational Awareness (SSA) scenarios with spawn events. 
\end{abstract}

\section{Introduction}
In this paper, we present a randomized approach to the multi-object tracking problem called randomized Finite Set Statistics (R-FISST). This approach allows us to approximate the full Finite Set Statistics (FISST) Bayesian recursions, account for multiple birth and death scenarios, while keeping the problem computationally tractable. We briefly show that the FISST recursions can essentially be considered as a discrete state space Bayesian filtering problem on ``Hypothesis Space" with the only input from the continuous problem coming in terms of the likelihood values of the different hypotheses. It is this perspective that allows us to develop the randomized approach where we sample the highly probable hypotheses using a Markov Chain Monte Carlo (MCMC) technique. Our previous papers, \cite{Faber1,Faber2}, used similar techniques to generate hypotheses, however, they required that all possible birth and death hypotheses be considered which limited us to cases with only one birth or death for computational tractability. The newly developed R-FISST can account for multiple births and death within the MCMC hypothesis generation and alleviates the computational burden of the  birth and death model. This increases the R-FISST method's ability to handle situations of mass birth and death, or spawning.  This new technique is tested using Space Situational Awareness (SSA) scenarios that include birth, death, and spawning. 

FISST-based multi-object detection and tracking has been developed based on the mathematical theory of finite set statistics \cite{Mahler:97,Mahler:07}.  The greatest challenge in implementing FISST in real-time, which is critical to any viable SSA solution, is computational burden. The first-moment approximation of FISST  is known as the Probability Hypothesis Density (PHD) approach \cite{Mahler:07,Vo06}. The PHD has been proposed as a computationally tractable approach to applying FISST. The PHD filter essentially finds the density of the probability of an object being at a given location, and thus, can provide information about the number of objects (integral of the PHD over the region of interest) and likely location of the objects (the peaks of the PHD). The PHD can further employ a Gaussian Mixture (GM) or a particle filter approximation to reduce the computational burden (by removing the need to discretize the state space). This comes at the expense of approximating the general FISST pdf with its first-moments \cite{Vo06, Vo_SMC, Vo_CPHD, Clark_PHD}. The PHD filter does not attempt to solve the full FISST recursions, in particular, by considering the PHD, the filter gets rid of the data association problem inherent in these problems. In other previous work, a GM approximation was applied, not to the first-moment of the FISST pdfs, but to the original full propagation and update equations derived from FISST \cite{Hussein:12GNC,Hussein:12b}. This eliminates any information loss associated with using the first-moment PHD approximation, while at the same time increasing the computational tractability of the multi-object FISST pdfs. More recently, there has been substantive research on the so-called generalized labeled multi-Bernoulli (GLMB) filters that model the birth process as a multi-Bernoulli process and tractable implementation of the filter have been proposed based on a lookahead strategy based on the cheaper PHD filter \cite{MB_FISST1, MB_FISST2}. Our birth and death model is also a multi-Bernoulli process, however, our implementation is at a hypothesis level: our pdf is a weighted mixture of MT-pdfs with varying number of targets in the component MT-pdfs. The weights are precisely the hypothesis weights and an MCMC scheme is used to sample the high probability hypothesis thereby resolving the problem of hypothesis depletion, especially when the number of hypotheses become very large, for instance in the case of a spawning event that is considered here (the number of hypothesis in such cases can explode to the billions even for a moderate number of objects).

%In this paper, in contrast, we introduce a hypothesis level derivation of the FISST equations that makes it clear as to how the full FISST recursions can be implemented without any approximation other than the approximations inherent in the underlying tracking filter, such as an extended  Kalman filter. We introduce a simplified model for the birth and death process that allows for only one birth or one death in any time interval, thereby controlling the number of birth and death hypotheses while still being able to converge to the correct hypothesis regarding the total number of objects given the birth objects remain in the field of view for long enough. Further, in order to ensure the computational tractability of the resulting equations, we introduce an MCMC based hypothesis selection scheme resulting in the Randomized FISST (R-FISST) approach that is able to scale the FISST recursions to large scale problems. We call our method R-FISST, since as in FISST, the hypotheses in our method have varying number of objects, and in essence, give a probabilistic description of the random finite set representing the multi-target probability distribution. We also formally show the equivalence of the two methods in the appendix.
Hypothesis Oriented MHT (HOMHT) \cite{HOMHT, BarShalom1, BarShalom2, mcmcda}, and the Track Oriented MHT (TOMHT) \cite{TOMHT} are non-FISST approaches to the multi-object tracking problem. The MHT techniques can be divided into single-scan and multi-scan methods depending on whether the method uses data from previous times to distinguish the tracks \cite{BarShalom1, jpda, mcmcda}. We show that  the MHT technique and the FISST technique are essentially the same, modulo the set-theoretic representation of multi-target pdfs in FISST (which, however, does not provide any extra information). This is made possible through the hypothesis level derivation of the tracking equations that considers the full hybrid state of the problem unlike the original MHT derivation in \cite{HOMHT}. This allows us to identify the critical structure inherent in the FISST recursions that enables us to unify the two approaches: essentially our approach results in a mixture of hypotheses, where the hypotheses themselves are MT-pdfs with possibly different number of objects. This also allows for the computationally tractable MCMC based implementation of the full FISST recursions even when the number of hypotheses grow extremely large.

The rest of the paper is organized as follows. In Section II, we give a brief introduction to the hypothesis level derivation of the FISST equations and show the relationship to MHT. In Section III, we show how to incorporate birth and death into the MCMC based randomized hypothesis selection technique that results in the newly developed RFISST algorithm. In Section IV, we show an application of the RFISST technique to multiple spawning SSA scenarios that highlight the necessity of the randomized approach.    %A related paper,\cite{Faber1}, was presented at the International Conference of Information Fusion. This paper extends that paper with a new MCMC data association scheme and a full comparison of the methodology with the HOMHT technique. For the sake of the paper being self-contained, the next two sections detailing the hypothesis level derivation of the FISST equations are reproduced from reference \cite{Faber2}

\section{A brief overview of the Hypothesis based Derivation of the FISST equations}
%In this section, we shall  frame the multi-object tracking equations at the discrete hypothesis level ( which we believe are equivalent to the FISST equations) which then shows clearly as to how the full FISST recursions may be implemented. The derivation below assumes that the number of measurements is always less than the number of objects, which is typically the case in the SSA problem.  We never explicitly  account for the number of objects, since given a hypothesis, the number of objects and their probability density functions (pdf) are fixed, which  allows us to derive the results without having  to consider the random finite set (RFS) theory underlying FISST.  Albeit the equations derived are not as general as the FISST equations, in particular, the birth and death models employed here are quite simple, we believe that the level of generality is sufficient for the SSA problem that is our application. 
In this section, we give a very breif overview of the hypothesis level derivations to the full FISST recursions. We will discuss only the key points of the derivation that allow us to develop the computational technique presented later in this paper. For more detailed derivations, please refer to \cite{Faber1,Faber2}.
\subsection{Framing FISST at the Hypothesis Level}
We consider first the case when the number of objects is fixed. Assume that the number of objects is $M$, and each object state resides in $\Re^N$. Consider some time instant $t-1$, and the data available for the multi-object tracking problem till the current time $\mathcal{F}^{t-1}$. Let $H_i$ denote the $i^{th}$ hypothesis at time $t-1$, and let $\{X\}$ denote the underlying continuous state. For instance, given the $N-$ object hypothesis, the underlying state space would be $\{X\} = \{X_1, X_2,\cdots X_N\}$ where $X_j$ denotes the state of the $j^{th}$ object under hypothesis $H_i$ and resides in $\Re^N$. Let $p(\{X\}, i/ \mathcal{F}^{t-1})$ denote the joint distribution of the state-hypothesis pair after time $t-1$. Using the rule of conditional probability:
\begin{align}
p(\{X\}, i/ \mathcal{F}^{t-1}) = \underbrace{p(\{X\}/ i, \mathcal{F}^{t-1})}_{\mbox{MT-pdf underlying} H_i} \underbrace{p(i/ \mathcal{F}^{t-1})
}_{w_i = \mbox{prob. of} H_i},
\end{align} 
where MT-pdf is the multi-object pdf underlying a hypothesis. Given the hypothesis, the MT-pdf is a product of independent individual pdfs underlying the objects, i.e.,
\begin{align}
p(\{X\}/ i, \mathcal{F}^{t-1}) = \prod_{k=1}^M p_k(x_k),
\end{align}
where $p_k(.)$ is the pdf of the $k^{th}$ object.
\begin{remark}
In random finite set theory, the arguments of the MT-pdf $\{x_1, x_2 \cdots x_M\}$ above are interchangeable and thus, the MT-pdf is represented as:
\begin{align}
p(\{X\}/ i, \mathcal{F}^{t-1}) = \sum_{\bar{\sigma}}\prod_{k=1}^M p_{\sigma_k}(x_k),
\end{align}
where $\bar{\sigma} = \{\sigma_1, \sigma_2 \dots \sigma_M\}$ represents all possible permutations of the indices $\{1,2 \cdots M\}$. Hence, in any integration involving such a set of indices, a normalizing factor of $\frac{1}{M!}$ is used. In our case, we explicitly assign the index $x_k$ to the target $k$ or more precisely, the $k^{th}$ component of the MT-pdf, $p_k(.)$. Note that such an assignment is always possible and there is no information loss in such a representation. Moreover, at the expense of  more bookkeeping, this allows us to keep track of the labels of the different components of our multi-target pdfs. Please see the reference \cite{HFISST_journal} where we show the equivalence of the hypothesis level equations derived here and the FISST recursions.
\end{remark} 

Next, we consider the prediction step between measurements. Each $M$-object hypothesis $H_i$ splits into $A_M$ children hypotheses
\begin{align}\label{n_da}
A_M = \sum_{n=0}^{min(m,M)} {M \choose n} {m \choose n} n!,
\end{align}
where $m$ is the number of measurement returns. We note here that this is a pseudo-prediction step since we assume that we know the size of the return $m$. However, it allows us to fit the MT-tracking method nicely into a typical filtering framework.
Using the rules of total and conditional probability, it follows that the predicted multi-object pdf in terms of the children hypotheses is:
\begin{align} \label{FISST_pred}
p^-&(\{X\}, (i,j)/ \mathcal{F}^{t-1}) = \nonumber\\
&\underbrace{\int p(\{X\}/(i,j), \{X'\})p(\{X'\}/ i, \mathcal{F}^{t-1}) d\{X'\}}_{p^-(\{X\}/ (i,j), \mathcal{F}^{t-1})}\nonumber\\
&\hspace{2ex}\underbrace{p(j/ i)}_{p_{ij}} 
\underbrace{p(i/ \mathcal{F}^{t-1})}_{w_i},
\end{align}
where $p^{-}(., (i,j)/ \mathcal{F}^{t-1})$ is the joint distribution of the state and hypothesis pairs before the measurement at time $t$, $p_{ij}$ is the transition probability of going from the parent $i$ to the child $j$ and $w_i$ is the probability of the parent hypothesis $H_i$.  Let $p_k(x_k/ x_k')$ denote the transition density function of the $k^{th}$ object. 
\begin{align}
p^-&(\{X\}/ (i,j), \mathcal{F}^{t-1}) \equiv \nonumber\\
&= \prod_ k \int p_k(x_k/ x_k') p_k(x_k') dx_k' = \prod_k p_k^-(x_k),
\end{align}
where $p_k^-(x_k)$ is the prediction of the $k^{th}$ object pdf underlying the hypothesis $H_{ij}$. 
%\begin{remark}
%Eq. \ref{FISST_pred} has a particularly nice hybrid structure: note that the first factor is the multi-object continuous pdf underlying the child hypothesis $H_{ij}$, while the second factor $p_{ij} w_i$ is the predicted weight of the hypothesis $H_{ij}$.
If a priori information, for instance, in terms of a probability of detection $p_D$ is available, then:
\begin{align} \label{hyp_prior_wt}
p_{ij} = \frac{p_D^k (1-p_D)^{M-k}}{{m\choose k} k!},
\end{align}
where $ij$ is a data association in which exactly $k$ of the $M$ targets are associated to measurements. The ${m\choose k} k!$ factor is required so that $p_{ij}$ is a valid probability distribution , i.e., $\sum_j p_{ij} = 1$. 
%\begin{align}
%\sum_j p_{ij} = \sum_{k=0}^M {M \choose k} \frac{p_D^k(1-p_D)^{M-k}}{{m\choose k}k!} {m\choose k}k! = 1.
%\end{align}
%Note that the MT-pdf underlying $H_{ij}$ is simply the product of the predicted individual object pdf, and in the case of no birth and death, it is the same for all children hypothesis $H_{ij}$. 
%\end{remark}

Given the prediction step above, let us consider the update step given the measurements $\{Z_t\} = \{z_{1,t},\cdots z_{m,t}\}$. 
Using Bayes rule:
\begin{align}\label{FISST_M}
&p(\{X\}, (i,j)/ \mathcal{F}^t) = \nonumber\\
&\frac{p(\{Z_t\}/ \{X\}, (i,j))p^-(\{X\}/ (i,j), \mathcal{F}^{t-1})}{l_{ij}}
\frac{l_{ij}\overbrace{p_{ij}w_i}^{w_{ij}}}{\sum_{i',j'} l_{i',j'} \underbrace{p_{i'j'}w_{i'}}_{w_{i'j'}}},
\end{align}
where
\begin{align}
l_{ij} = \int p(\{Z_t\}/ \{X'\}, (i,j))p^-(\{X'\}/ (i,j), \mathcal{F}^{t-1}) d\{X'\}.
\end{align}
Note that $l_{ij}$ is likelihood of the data $\{Z_t\}$ given the multi-object pdf underlying hypothesis $H_{ij}$, and the particular data association that is encoded in the hypothesis, i.e., $z_{i}\mapsto{x_{ji}}$. Then,
\begin{align}\label{hyp_likelihood}
l_{ij} = \prod_{i=1}^m p(z_i/ X_{j_i}),
\end{align}
where
\begin{align}
p(z_i/ X_{j_i}) = 
\begin{cases}
    \int p(z_i/ x) p_{j_i}(x) dx & \text{if } X_{j_i} \in \mathcal{T}\\
    g(z_i)  & \text{if }  X_{j_i} \in \mathcal{C} 
    \end{cases}
\end{align}
where $\mathcal{T}$ is the set of all objects and $\mathcal{C}$ is clutter and the above equation implies that the measurement $z_i$ was associated to clutter if $X_{j_i} \in \mathcal{C}$.
%\end{remark}
Thus, given that the likelihoods of different hypothesis $l_{ij}$ arise from the underlying multi-object pdf and the encoded data association in the hypotheses $H_{ij}$, the FISST updates can be written purely at the hypothesis level as follows:
\begin{align}\label{FISST_M'}
w_{ij} := \frac{l_{ij}w_{ij}}{\sum_{i',j'} l_{i'j'}w_{i'j'}},
\end{align}
where $w_{ij} = p_{ij}w_i$. Thus, we can see that the FISST update has a particularly simple Bayesian recursive form when viewed at the discrete hypothesis level, given that the multi-object pdfs underlying the hypotheses $H_{ij}$ are tracked using some suitable method. 
The hypotheses $H_{ij}$ now become the parent hypotheses for the next time step.

\subsection{Relationship to MHT} 
 The MHT likelihood for a child hypothesis $j$ of parent $i$ is of the form:
\begin{align}
\eta_{ij}^{MHT}(z_1,..z_m) = p_D^k (1-p_D)^{M-k} \prod_{l=1}^m p(z_l/ \hat{x}_{j_l}),
\end{align}
where $\hat{x}_{j_l}$ is the mean of the pdf of the target $X_{j_l}$ or $p(z_k/ \hat{x}_{j_l}) = g(z_l)$ if measurement  $z_l$ is associated to clutter. The equivalent hypothesis likelihood in the hypothesis level FISST (H-FISST) derivation is:
\begin{align}
&\eta_{ij}^{HFISST}(z_1,..z_m) = \nonumber\\
&p_{ij} l_{ij}=  \frac{{p_D}^k (1-p_D)^{M-k}}{{m \choose k}k!}\prod_{l=1}^m p(z_l/ X_{j_l}),
\end{align}
where the terms under the product in the above equation have been defined in Eq. \ref{hyp_likelihood}. Thus, it may be seen that the main difference in the two likelihoods is the factor ${m \choose k} k!$ and the fact that $p(z_l/ \hat{x}_{j_l})$ is an approximation of $p(z_l/ X_{j_l})$ for observations that have been associated to a target. The factor is required such that the likelihood is normalized. The normalization of the likelihood, as in HFISST, is necessary from a probabilistic perspective since otherwise the distribution on the filtered state pdf, resulting from all possible observations, does not constitute a probability distribution, i.e., it does not add up to unity. This can easily be seen for the case of a standard filtering problem which carries over to the mutli-target tracking problem. Let the filtered state pdf, the belief state, be denoted by $b(x)$. Suppose that the likelihood function $\int p(z/x) dz \neq 1$. Consider the distribution on the future belief state $b'(x)$. This is given by:
\begin{align}
p(b'/b) = \int_z p(b'/z,b) p(z/b) dz, \mbox{where} \nonumber\\
p(z/b) = \int p(z/x) b(x) dx.
\end{align}
Note that if $\int p(z/x)dz \neq 1$ then $\int p(z/b) dz \neq 1$. Hence,
\begin{align}
\int p(b'/b) db' = \int\int p(b'/z,b) p(z/b) dz db' \nonumber\\
=  \int p(z/b) dz \neq 1.
\end{align}
We know that the filtered pdf (the belief process) has to evolve according to a Markov chain \cite{bertsekas1, kumar1} but the above development shows that the evolution equation violates the requirement that the transition probability of a Markov chain needs to be a probability distribution over all future states,  if the likelihood does not normalize to unity.

\subsection{Equivalence of MHT and FISST}
The proposed hybrid derivation (in that it includes both the continuous and discrete parts of the problem) as opposed to MHT which is a purely discrete derivation at the hypothesis level \cite{HOMHT},  reveals the critical hybrid structure (Eq. \ref{FISST_M}) inherent to  multi-target tracking problems, and which, in turn  allows us to unify the HFISST development with the FISST based formulation of the multi-target tracking problem, and thus, allows for the unification of FISST and MHT: methods that have thus far been thought to be different from each other. Because of the paucity of space, we cannot reproduce the derivation here, but the reader is referred to the technical report \cite{HFISST_journal} for more details.

\section{A Randomized FISST (R-FISST) Technique}
In this section, we show how birth and death processes can be conveniently included in the HFISST technique, and propose a randomized implementation of the same, termed Randomized FISST (RFISST).

\subsection{Incorporating Birth and Death Processes into HFISST}
\begin{figure}[h]
\centering
\includegraphics[scale=.30]{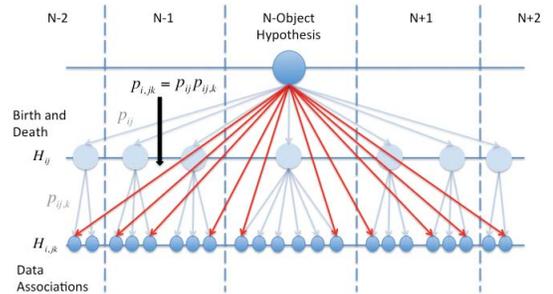}
\caption{An illustration of the multiple birth and death process. Arrows show connection between parent, child, and grandchild hypotheses.}
\label{tree}
\end{figure}

We assume the following model for the brith and death of targets. 
\paragraph{Birth and Death Process:} We assume that the births and deaths can only happen in the field of view (FOV) of the sensor. This is done to ensure computational tractability but can be relaxed for theoretical purposes. We further assume that the FOV has been discretized into $N$ pixels. Further, let us consider an $M$-target hypothesis $H_i$. We assume that the birth process is a simple Binomial process where the probability of a birth in a given pixel at a given time instant is $\alpha$, and this is independent of the birth in any other pixel. Further, we  assume that the death process is also Binomial, and any target in the FOV can disappear with a probability $\beta$ independent of the other targets in the FOV. Moreover, the birth and death processes are independent of each other. \\
If we let the number of pixels tend to infinity, and we let the number of time instants, $n$, go to infinity, such that $n\alpha = \lambda$, we recover a Poisson Point process.\\
It may be seen that the probability of any instance of exactly $N_b$ births and $N_d$ deaths starting with the M-target hypothesis given by:
\begin{align}
p_{ij} = \alpha^{N_b}\beta^{N_d},
\end{align}
and the number of such instances is ${N\choose {N_b}} {M\choose {N_d}}$. Given the child $H_{ij}$ of the parent $H_i$ that incorporates a particular number of birth and death hypothesis, due to the data association, the child $H_{ij}$ can further split into grandchildren $H_{ijk}$ where, from before, the probability of the $k^{th}$ data association hypothesis given the $ij^{th}$ child is:
\begin{align}
p_{ij, k} = \frac{p_D^l (1-p_D)^{m-l}}{{m\choose l} l!},
\end{align}
where $m$ is the number of returns, and the $ijk^{th}$ child corresponds to a data association hypothesis that chooses to associate $l$ of the returns to objects in the FOV. 
Thus, given a particular $M$ object hypothesis, the total number of possible hypotheses (at the grandchild level) after receiving a measurement is:
\begin{align}\label{n_da_m}
\widetilde{A}_{M} = \sum_{K=0}^{N} {a_{M+K}} {A_{M+K}} + \sum_{K=-1}^M a_{M-K} A_{M-K}, 
\end{align}
where $A_{M+K}$ is calculated using Eq. \ref{n_da} and,
\begin{align}\label{n_da_a}
a_{M+K}= \sum_{j=0}^{N} {N \choose {K+j}} {M \choose {j}},\\
a_{M-K} = \sum_{j=0}^{M-K} {M \choose {K+j}} {N \choose j}.
\end{align}
The birth hypothesis includes a continuous pdf for any birthed object to form the MT-pdf underlying the birth hypothesis, we do not go into the details here but this may be done in a straightforward, albeit tedious, fashion and is shown in our previous work \cite{Faber1, Faber2}. Thus, the probability of the $jk^{th}$ child given the parent hypothesis $H_i$ is simply:
\begin{align} \label{Tr_bd}
p_{i,jk} = p_{ij}p_{ij,k}.
\end{align}
The likelihoods of any of these hypothesis, $l_{ijk}$, can now be calculated using Eq. \ref{hyp_likelihood} from before. Noting that $jk$ can be replaced by a single number $l$, it follows that the HFISST procedure carries through to the birth/ death case with minimal changes, except now the transition probability for a particular (grand)child hypothesis is given by Eq. \ref{Tr_bd}. However, note that the number of (grand)children hypothesis explodes further due to the birth and death process. Please refer to Fig. \ref{tree} for an illustration of the process. 
%$N$ is the number of pixels or discrete grids in the field of view and is a result of our model for the arrivals of birth and death. That being said our model for birth and death consists of discritizing the field of view. In each discrete grid we say the probability of having a birth at that location is $\alpha$ the probability of a death is $\beta$. Due to the structure of the HFISST derivation the only other neccessary change is of the hypothesis probabilities $p_{ij}$. For multiple birth and death with a priori sensor information the equation for the probability of the child hypothesis becomes,
%\begin{align}\label{pij_m}
%p_{ij}= p(z_k/X_{jk})p_D^k(1-p_D)^{M-k}\\
%\times{ \ g(z)p_f^{m-k-N_{\alpha}}(1-p_f)^{k+N_{\alpha}}}\\
%\times{ \  \alpha^{N_{\alpha}}\beta^{N_\beta}},
%\end{align}
%where $N_\alpha$ and $N_\beta$ are the number of births and deaths respectively. 
%With these changes we can sample the child hypotheses using an MCMC approach that favors the probable hypotheses. This gives us the ability to generate the highly probable hypotheses without having to enumerate all $\widetilde{A_M}$ possible hypotheses. 
\subsection{MCMC based Intelligent Sampling of Children Hypothesis}
Recall Eq. \ref{FISST_M'}. It is practically plausible that most children $j$ of hypothesis $H_i$ are highly unlikely, i.e., $l_{ij} \approx 0$ and thus, $w_{ij} \approx 0$. Hence, there is a need to sample the children $H_{ij}$ of hypothesis $H_i$ such that only the highly likely hypotheses are sampled, i.e., $l_{ij} >> 0$. 

\begin{figure}[h]
\centering
\includegraphics[scale=.30]{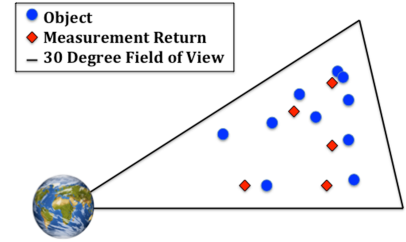}
\caption{A possible SSA event where there is assumed to be ten objects in the field of view and five measurement returns.}
\label{EXsen}
\end{figure}

\begin{figure}[h]
\centering
\includegraphics[scale=.30]{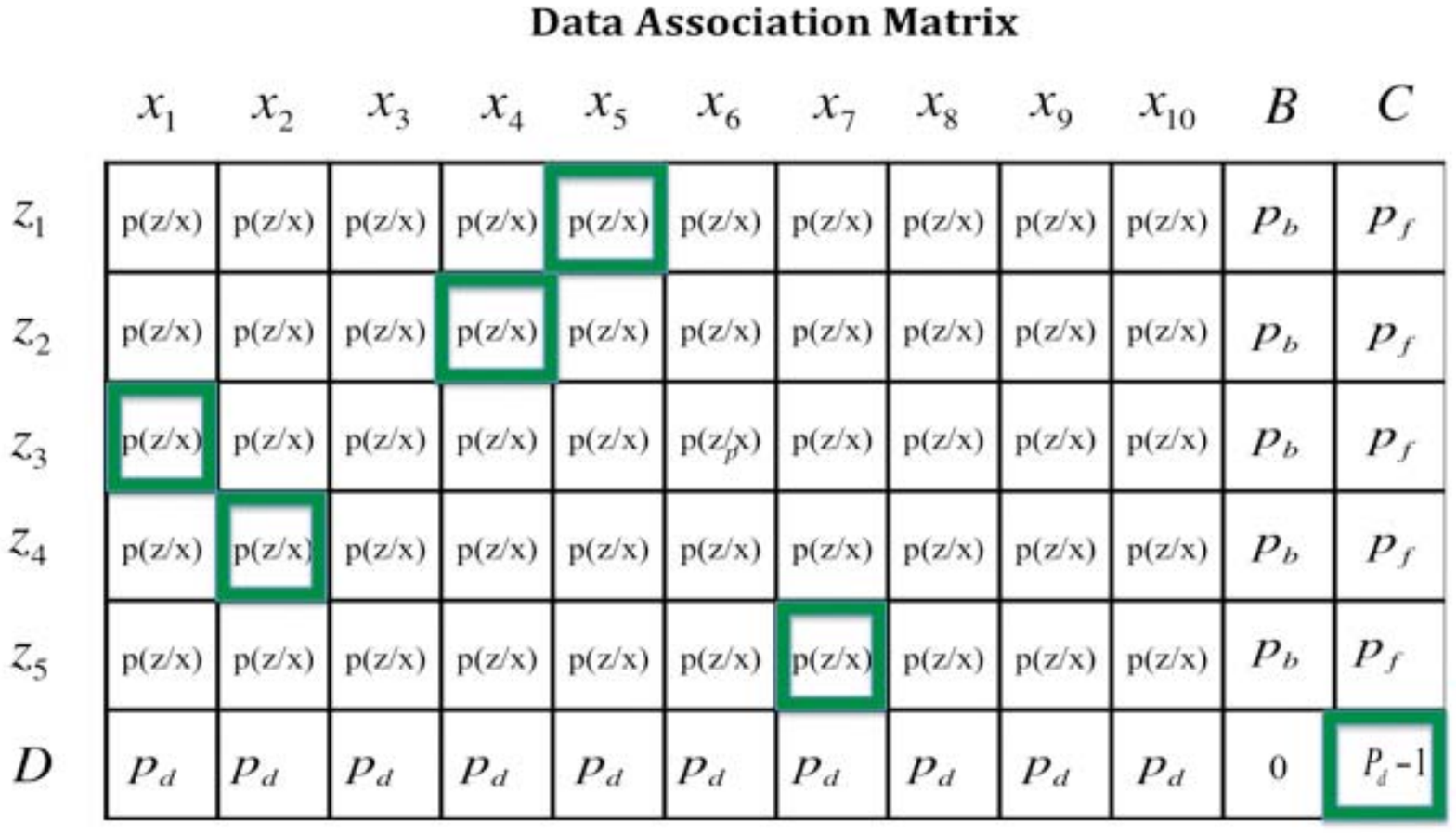}
\caption{The Data Association Matrix. Each row represents a particular measurement return. Each column is a particular association. The elements represent the likelihood of the corresponding association. Green boxes here show a visual representation of an example hypothesis.}
\label{da}
\end{figure}

\begin{remark}
Searching through the space of all possibly hypotheses quickly becomes intractable as the number of objects and measurements increase, and int he presence of multiple birth and deaths.
We cannot sample the hypothesis naively  either, for instance, according to a uniform distribution, since the highly likely hypothesis are very rare under the uniform distribution, and thus, our probability of sampling a likely hypothesis is vanishingly small under a uniform sampling distribution. This is the equivalent to the "particle depletion" problem seen in particle filters.
\end{remark}
Thus, we have to resort to an intelligent sampling technique.
In particular, given a hypothesis $H_i$, we want to sample its children according to the probabilities $\bar{p}_{ij} = w_{ij}l_{ij}$. This can be done by generating an MCMC simulation where the sampling Markov chain, after enough time has passed (the burn in period), will sample the children hypotheses according to the probabilities $\bar{p}_{ij}$.  A pseudo-code for setting up such an MCMC simulation is shown in Algorithm 1.
\begin{algorithm}
\caption{MCMC Hypothesis Sampling}
Generate child hypothesis $j_0$, set $k=0$.\\
Generate $j_{k+1} = \pi(j_k)$ where $\pi(.)$ is a symmetric proposal distribution\\
If $\bar{p}_{ij_{k+1}}> \bar{p}_{ij_k}$ then $ j_{k} := j_{k+1}; k:= k+1$;\\
else $j_k := j_{k+1}$ with probability proportional to $\frac{\bar{p}_{ij_{k+1}}}{\bar{p}_{ij_k}}$; $k = k+1$.
\end{algorithm}
In the limit, as $k\rightarrow \infty$, the sequence $\{j_k\}$ generated by the MCMC procedure above would sample the children hypotheses according to the probabilities $\bar{p}_{ij}$. %Suppose that we generate $C$ highest likely distinct children hypothesis $H_{ij}$ using the MCMC procedure, then the FISST recursion Eq. \ref{FISST_M'} reduces to:
%\begin{align}
%w_{ij} := \frac{l_{ij}w_{ij}}{\sum_{i',j'} l_{i'j'}w_{i'j'}}, 
%\end{align}
%where $i'$ and $j'$ now vary from 1 to $C$ for every hypothesis $H_i$, instead of the combinatorial number $A_M$. 
Let us keep the highest likely $C$ children hypothesis for every hypothesis.
Given these $M*C$ hypotheses, i.e. $C$ children of $M$ parents, we can keep a fixed number $H_{\infty}$ at every generation by either sampling the $H_{\infty}$ highest weighted hypotheses among the children, or randomly sampling $H_{\infty}$ hypotheses from all the children hypotheses according to the probabilities $w_{ij}$. \\
%\begin{remark}
%The search for the highly likely hypotheses among a very (combinatorially) large number of options is a combinatorial search problem for which MCMC methods are particularly well suited. Thus, it is only natural that we use MCMC to search through the children hypotheses.
%\end{remark}
%\par
%\begin{remark}
%The choice of the proposal distribution $\pi(.)$ is key to the practical success of the randomized sampling scheme. Thus, an intelligent proposal choice is required for reducing the search space of the MCMC algorithm. We show such an intelligent choice for the proposal in the next section.
%\end{remark}
%\par
%\begin{remark}
%The discrete hypothesis level update Eq. \ref{FISST_M'} is key to formulating the MCMC based sampling scheme, and, hence, the computational efficiency of the R-FISST algorithm.
%\end{remark} 
%%%%%%%%%%%%%%%%%%%%%%%%%%%%%%%%%%%%%%%%%%%%%%%%%%%%%%%%%%%%%%%%%%%%%%%%%%%%%%%%%%
% WES sections
%\subsection{Smart Sampling Markov Chain Monte Carlo}
%In this section, we reveal the process used to perform the MCMC sampling discussed in the previous section. This process is performed at every scan to generate the highly likely children hypotheses. Consider the following SSA scenario depicted in figure \ref{EXsen}. In this scenario the belief is that there are ten objects in the field of view. The sensor then detects five measurement returns.
\begin{remark}
It may be shown that the mixing time of an MCMC chain is $\mathcal{O}(log N)$ where $N$ is the number of states in the chain \cite{MCMT} if the Markov Chain's  "congestion" is suitably bounded. The MCMC simulation below has this property, and thus, can scale to situations with a very large number of hypotheses. Due to the paucity of space, we do not show this here but this will be the subject of a forthcoming expanded journal version of the current paper.
\end{remark}
Now, we show the details of how to implement the multiple birth and death randomized hypothesis generation technique. Like our previously published papers, \cite{Faber1,Faber2}, this method takes advantage of MCMC at every time step to generate only the highly probable hypotheses. This method differs from previous methods because it accounts for multiple birth and death scenarios at each time step. Consider the scenario shown in figure \ref{EXsen}. In this scenario, a single hypothesis predicts there are ten objects in the field of view, which are represented by circles. A single sensor with a thirty-degree field of view takes a measurement and receives five returns represented by diamonds. Knowing the number of objects $M=10$ and the number of measurements $m=5$ the total number of possible hypotheses can be calculated using Eq. \eqref{n_da}, $A_{M}=63,591$.
%Typically when generating the hypotheses exhaustively one would create a matrix where each row represents a particular hypothesis. The columns of the matrix represent the measurement returns provided by the sensor. Each column entry represents the object that measurement is being associated to in the particular hypothesis. The hypothesis matrix for our example scenario would look like figure \ref{hm}. 
If we were to exhaustively generate the possible hypotheses from this particular scenario we would start by constructing a matrix called the hypothesis matrix. Each row of this matrix would represent a unique hypothesis. %Within the row, each column represents a particular measurement return while the elements represent the appropriate object association. An example matrix of this type can be seen in figure \ref{hm}. This matrix is similar to the one needed for the scenario depicted in figure \ref{EXsen} except for the fact that it would have $63,591$ rows.
%\begin{figure}[h]
%\centering
%\includegraphics[scale=.30]{HMb}
%\caption{An example of a typical Hypotheses Matrix used when exhaustively generating hypotheses. This particular matrix represents a portion of the hypothesis matrix that would be generated for the scenario in figure \ref{EXsen}.}
%\label{hm}
%\end{figure}
Generating this hypothesis matrix would become increasingly difficult as the number of objects and measurement returns increased. Even for a simple sixty object example with one spawn occurrence the maximum number of possible hypotheses at one time step can be in the order of tens of billions. Our randomized approach allows us to avoid generating this matrix in its entirety. Instead we only generate the rows of the hypothesis matrix that correspond to highly probable hypotheses. By taking each row of the hypothesis matrix as the state of a Markov Chain and we explore the chain using an MCMC criterion that favors the highly probable states. 
We can do this tractably by generating a matrix called the data association matrix (figure \ref{da}). Creating the data association matrix has no added computational cost because each element of the matrix is necessary in calculating hypothesis probability and is used in every method of hypothesis generation. Also, the dimensions of the data association matrix are much smaller than those of the hypothesis matrix. This makes it much more practical to explore using MCMC.  
%However, if all objects and measurements within the field of view can be associated then according to Eq. \eqref{n_da}, with $m=5$ and $M=10$, the total number of possible hypotheses would be $A_{M}=63,591$. Thus, the hypothesis matrix actually has $63,591$ rows. This illustrates the importance of a randomized approach. One can see that even with relatively low numbers of objects and measurement returns exhaustively generating and maintaining the hypotheses will cause a large computational burden. In our randomized approach we sample the rows of the hypothesis matrix based on hypothesis probability. We do this by creating a matrix we call the data association matrix, figure \ref{da}. 
%The data association matrix lists the objects as the columns and the measurement return as the rows. The entries of the matrix contain the likelihood value of that particular measurement to object assignment. The last column of the matrix is dedicated to clutter and contains the likelihood that a particular measurement is associated to clutter. The dimensions of this matrix are $m\times{(M+1)}$ which is much smaller than the dimensions of the hypothesis matrix. This makes it much more practical to explore using the MCMC technique. 
The data association matrix for the new randomized multiple birth and death approach has dimensions $(m+1)\times{(M+1+1)}$. That being said each row of the matrix represents a measurement return plus an added row to represent death. The columns represent the possible associations including a column for each associable object, a column for association to birth ($B$), and a column for no association ($C$). Each element of the matrix is the corresponding measurement to association likelihood.
\begin{remark}
The numbering of the objects and measurement returns in the data association matrix is done strictly for organization and is redone at random each time step with no record of previous numbering or labeling kept throughout scans. 
\end{remark}
%\begin{remark}
%Computing the data association matrix does not add any computational burden because the object to measurement likelihood is necessary in every tracking method.
%\end{remark}

%We start the randomized technique by creating a row vector of length $m$ containing a permutation of column numbers. The green boxes in figure \ref{da} are a visual representation of such a row vector $\left[ \begin{smallmatrix} 5&4&2&1&7 \end{smallmatrix} \right]$. This row vector is used as our first hypothesis. We then take a step in the MCMC by generating a proposed hypothesis. This is done by randomly choosing a row (Measurement) of the data association matrix to change. We then randomly sample a column (object) to associate the measurement to. If there is a conflicting assignment (i.e. a measurement is already assigned to that object) then we automatically assign the conflicting measurement to clutter.

\begin{figure}[h]
\centering
\subfigure[Choose a random return and association to switch]{
\includegraphics[scale=.25]{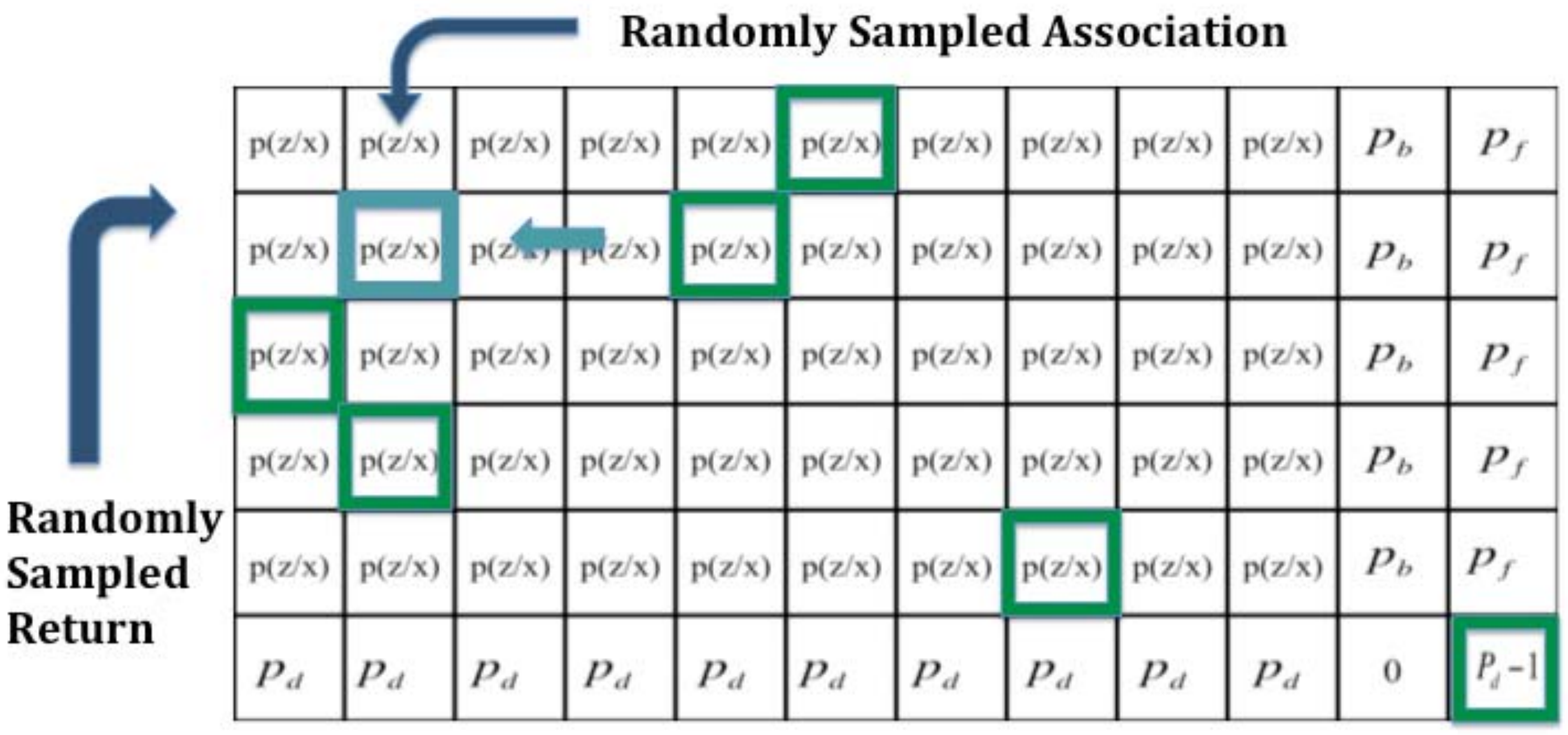}
\label{da2s1}
}
\subfigure[Check for conflicting assignment. Assign conflict to clutter if necessary]{
\includegraphics[scale=.25]{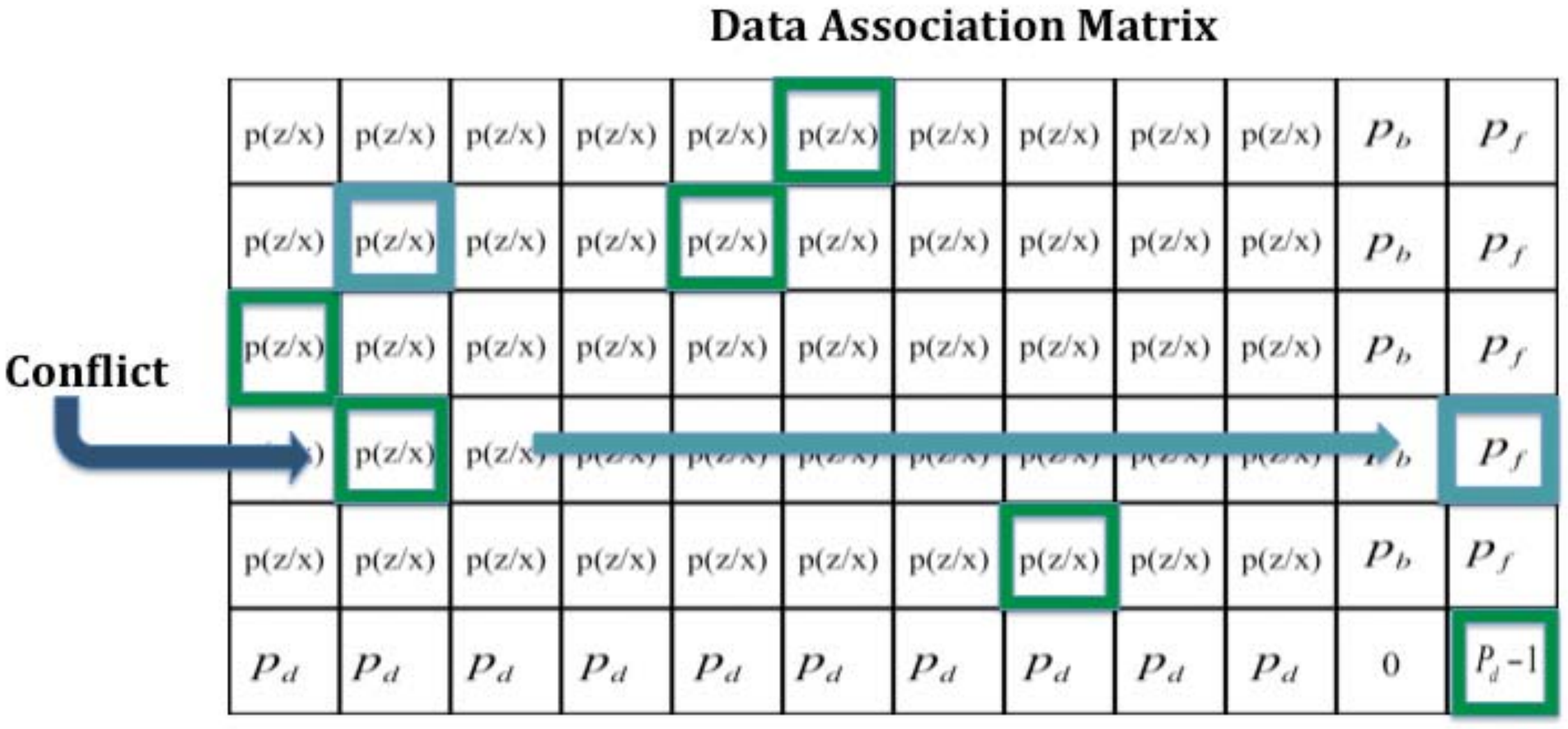}
\label{da2s1}
}
\subfigure[New proposed hypothesis]{
\includegraphics[scale=.25]{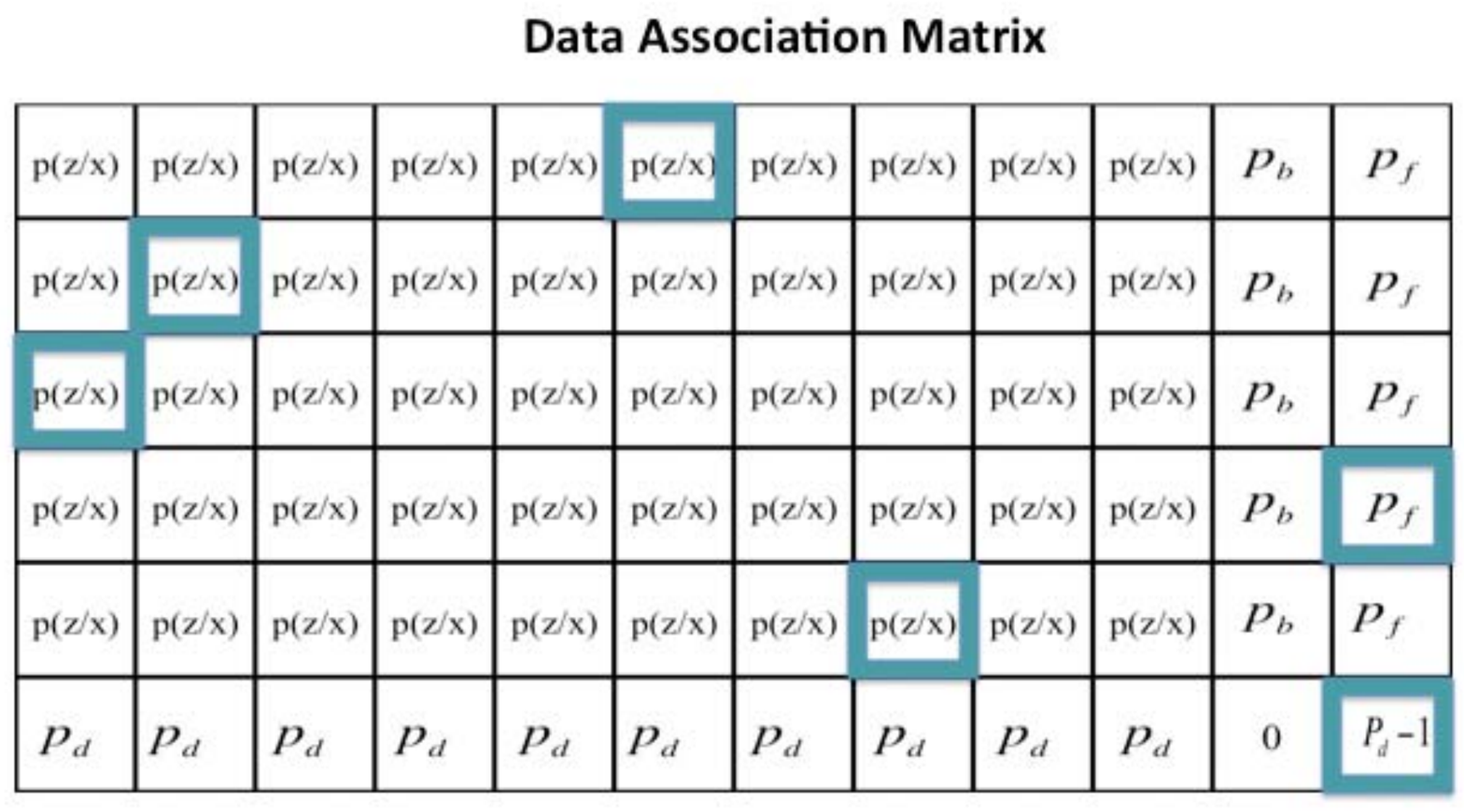}
\label{da2s1}
}
\caption{Visualization of a single MCMC step using the Data Association Matrix. This particular example contains a conflicting assignment with measurement return two and shows how the association is then changed to clutter.}
\label{da2}
\end{figure}
We start the MCMC procedure with a single row vector of length $m+1$ containing a random permutation of the numbers $1$ through $(M+2)$ and the letter N to signify a no death hypothesis. For example consider such a row vector $\left[ \begin{smallmatrix} 5&4&1&2&7&N \end{smallmatrix} \right]$.This row vector is our current hypothesis and is represented on the data association matrix in figure \ref{da} using boxes. We then propose a new hypothesis to walk to in the MCMC. We do this by randomly choosing a measurement return and switch its association using a uniform proposal distribution. We make sure the proposed switch creates a valid hypothesis with no conflicting associations. If there is a conflicting association we automatically assign the conflict to clutter. The resulting hypothesis becomes our proposed hypothesis. These steps are displayed in figure \ref{da2}.
\begin{remark}
Conflicting associations can be seen as more than one box in a single column. Columns $B$ and $C$ are an exception to this rule because there can be multiple associations to birth as well as clutter in a single hypothesis. Multiple birth hypotheses are developed naturally in the MCMC when more than one measurement to birth association probability is higher than or comparitive to the corresponding object associations. 
\end{remark}
We compare the proposed hypothesis to the current hypothesis using an MCMC criterion that stems from the Metropolis condition $U[0,1]< min( 1, \frac{\bar{p}_{ij_{k+1}}}{\bar{p}_{ij_{k}}})$ where $\bar{p}_{ij_{k}}$ is the probability of the hypothesis at step $k$. Simply, if the proposed hypothesis has a higher probability then it becomes the current hypothesis, if not, it becomes the current hypothesis with probability proportional to the ratio of the hypothesis probabilities. These steps are then repeated until we have reached an assumed stationary distribution. We then continue walking for a user defined amount of steps and record all the repeated hypotheses. The recorded hypotheses are those with very high probability.
\section{Applications}
%This section illustrates the application of the results from the previous sections. We illustrate the R-FISST based approach to the multi-object tracking and detection problem inherent in SSA applications. In particular, we will discuss the results from a fifty-space object birth and death scenario. Our goal is to show that the aforementioned methodology allows for accurate estimation while determining the correct number of objects in an environment where the number of objects is not fixed. This will allow for the methodology to be used in both catalog update and catalog maintenance. %More in depth results including comparrissons to other tracking methods such as HOMHT are to be presented in the 
In this section we apply the RFISST technique to a multi-object Space Situational Awareness (SSA) tracking problem. This simulation includes the birth of objects in the form of spawning to show how the method performs when there are multiple ambiguous measurement returns. Our goal is to highlight areas where the R-FISST approach is not only novel but neccessary to achieve accurate tracking data. We provide results from a twenty object Space Situational Awareness problem with a single spawn occurrence. We also show the growth of the number of possible hypotheses using a larger sixty object SSA example.

\subsection{R-Fisst Application to Multi-Object Tracking Scenario}
\begin{figure}[h]
\centering
\includegraphics[scale=.30]{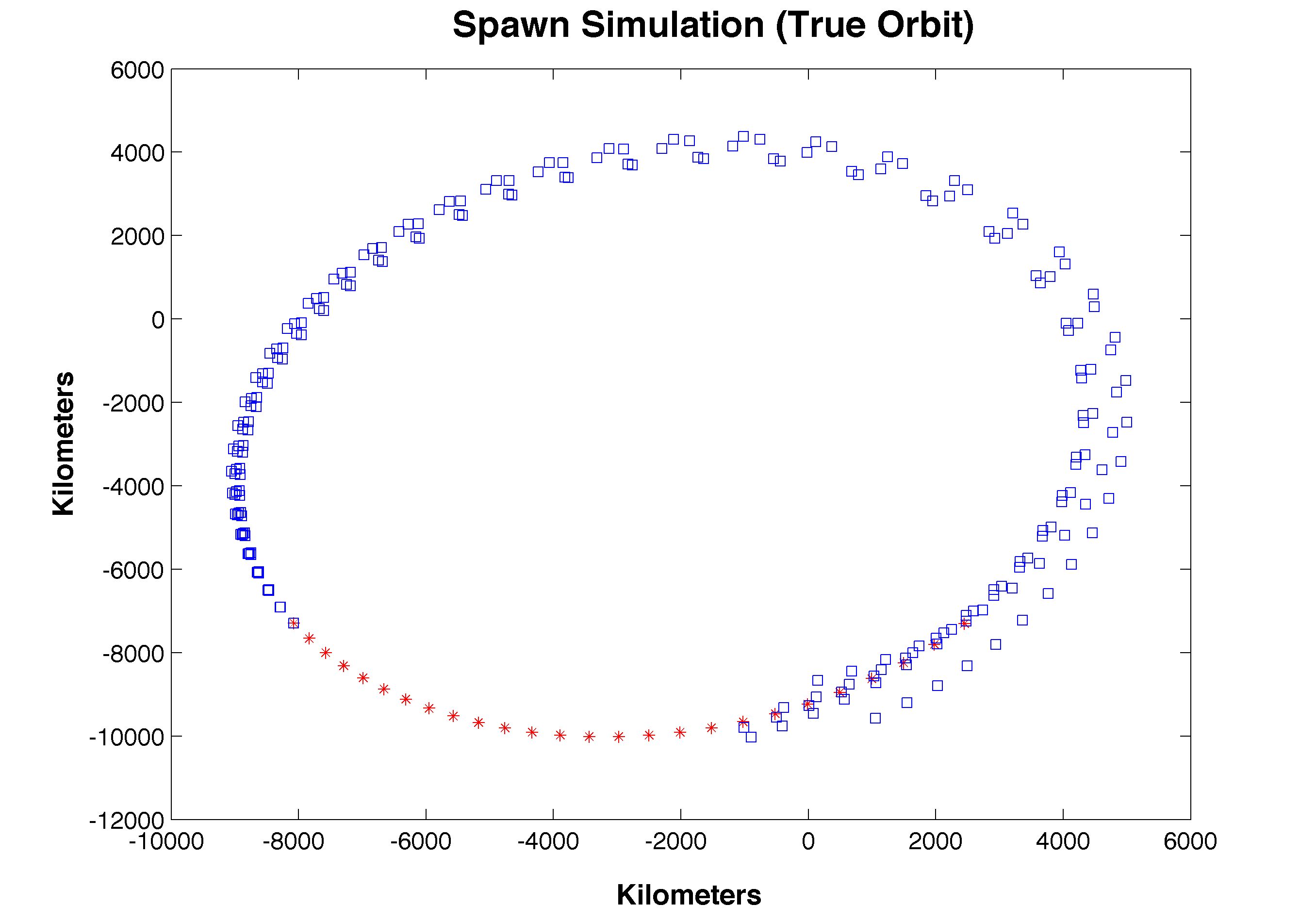}
\caption{Spawn simulation. A known object marked as red stars randomly splits into multiple objects (shown here as blue squares)}
\label{Spawn}
\end{figure}

We will test the algorithm developed in this paper by applying it to the SSA multi-object tracking problem. To show an instance were the randomized approach will be most useful we developed scenarios in which spawning occurs. We consider a spawn to be an event in which a known space object breaks apart into multiple objects. An example of such event is shown if figure \ref{Spawn}.

\begin{figure}[H]
\centering
\subfigure[Completely in the field of view.]{
\includegraphics[scale=.3]{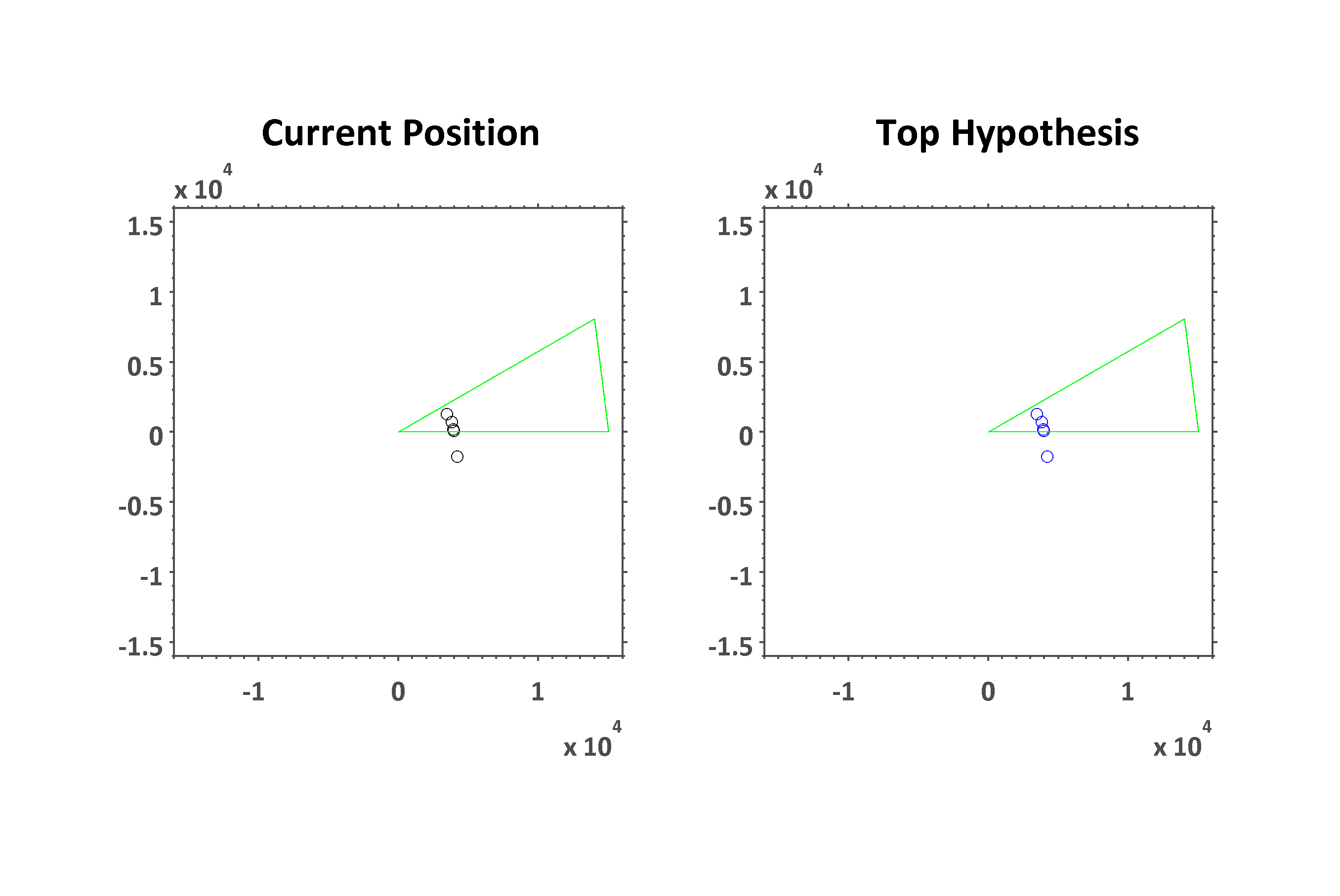}
\label{rs6}
}

\subfigure[After passing through the field of view.]{
\includegraphics[scale=.3]{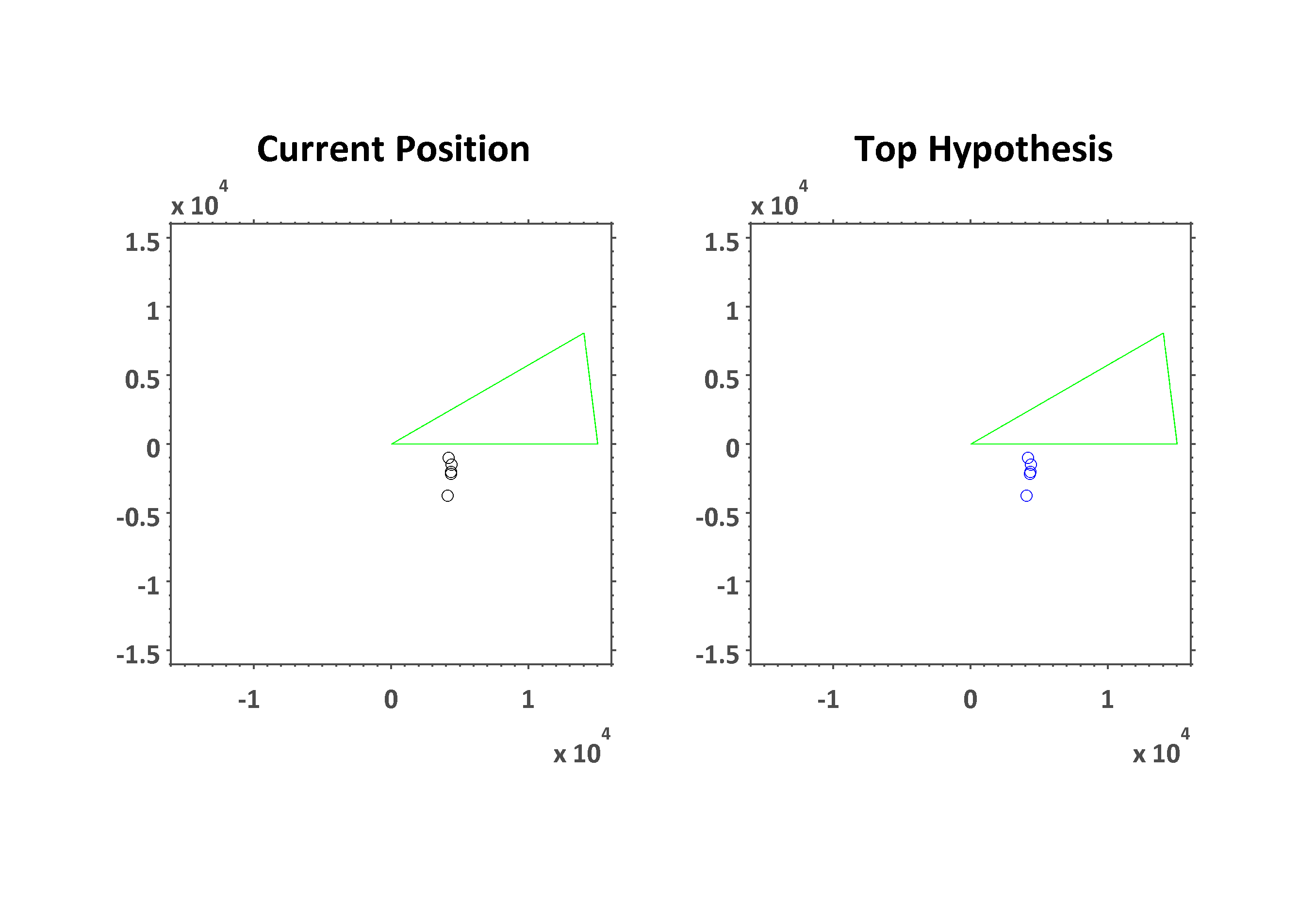}
\label{rs10}
}

\caption{Snapshots of the actual states (black) and the estimated states from the top hypotheses (green) while passing through the field of view. Axes in tens of thousands of kilometers}
\label{rs}
\end{figure}

These spawn events are of particular interest to our research because they create a scenario with multiple ambiguous measurement returns and lead to large numbers of birth and death hypotheses. This causes a rise in the total number of possible hypotheses. As a proof of concept we created a simple SSA tracking problem of a single object depicted in figure \ref{rs}. The initial state of this object was known with some uncertainty. A single sensor with a field of view of thirty degrees was used to capture measurements of the objects $x$ and $y$ position. At a random time in the orbit the object undergoes a spawn event. The goal of the simulation was to accurately estimate the states of all the newly spawned objects. Figure \ref{rs} shows the actual states of the spawned objects as they pass through the field of view next to the estimated positions from the top hypothesis. It can be seen from the figures that the R-FISST methodology was able to accurately predict the birth and track all the spawned objects. It is important to note that the method also correctly predicted that the original object had died. If it did not predict a death, the hypothesis would contain the wrong number of objects.

\begin{figure}[H]
\centering
\subfigure[Before measurements.]{
\includegraphics[scale=.3]{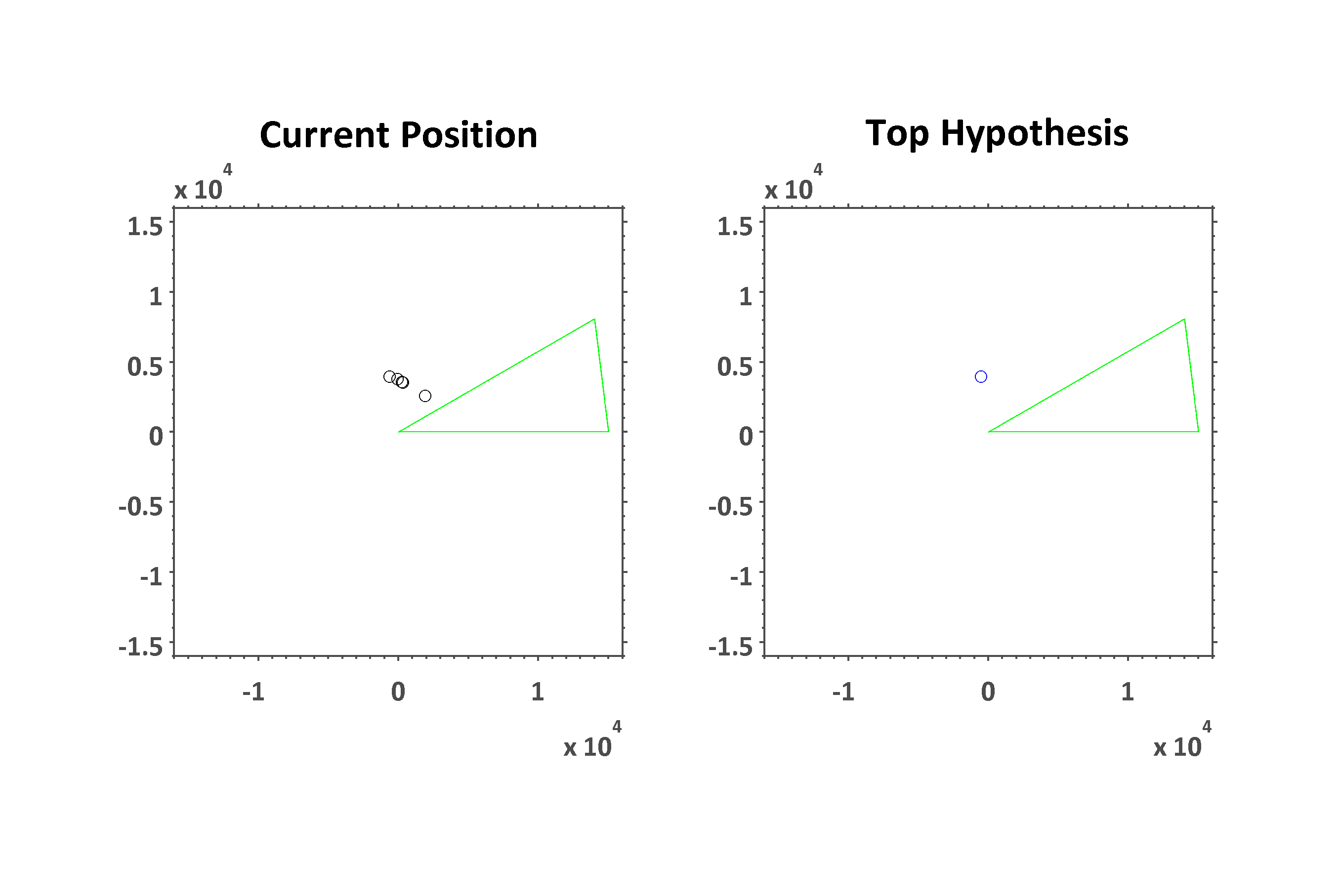}
\label{rs2}
}

\subfigure[Partially in the field of view.]{
\includegraphics[scale=.3]{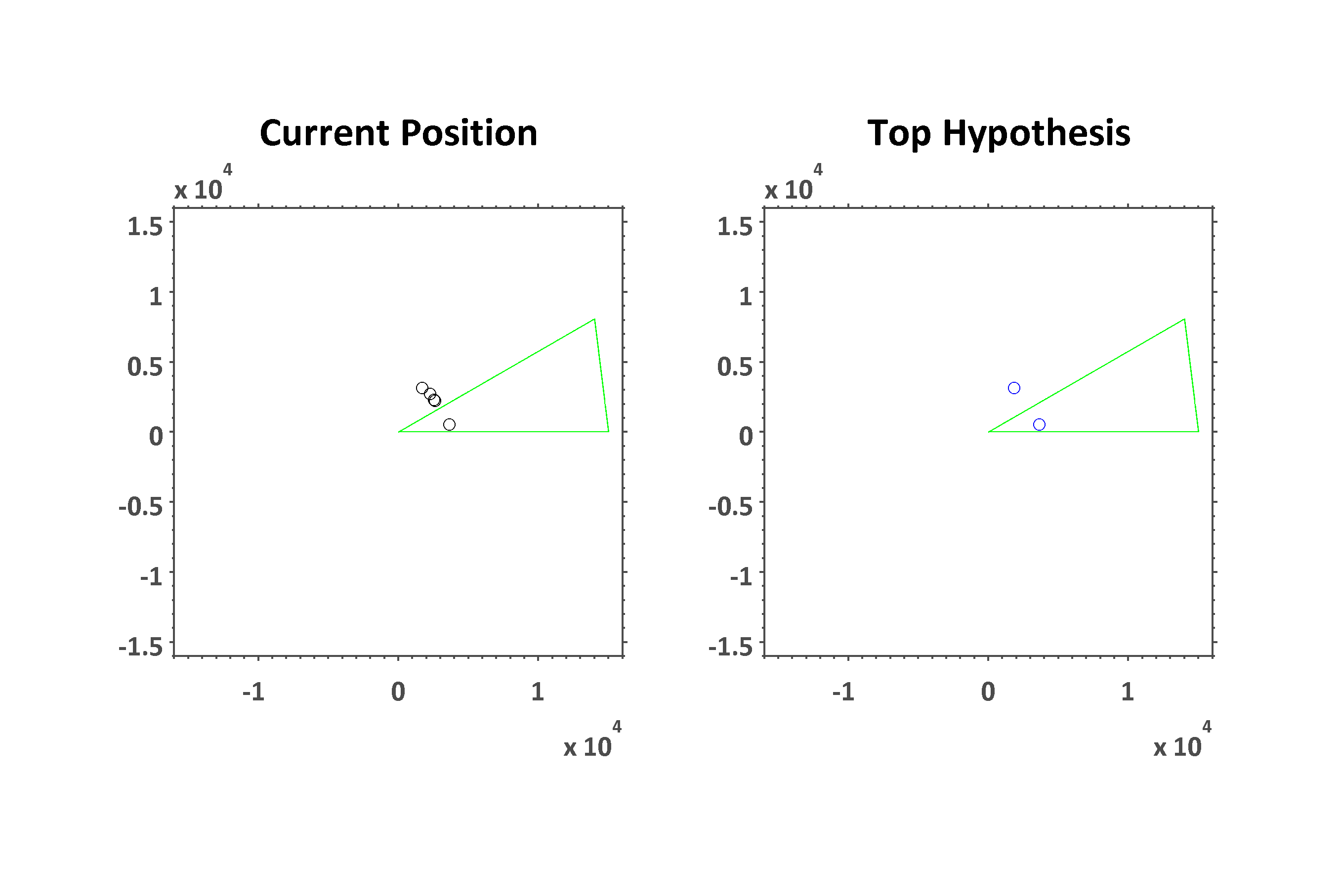}
\label{rs4}
}
\end{figure}

\begin{figure}[H]
\centering
\subfigure[Before measurements.]{
\includegraphics[scale=.07]{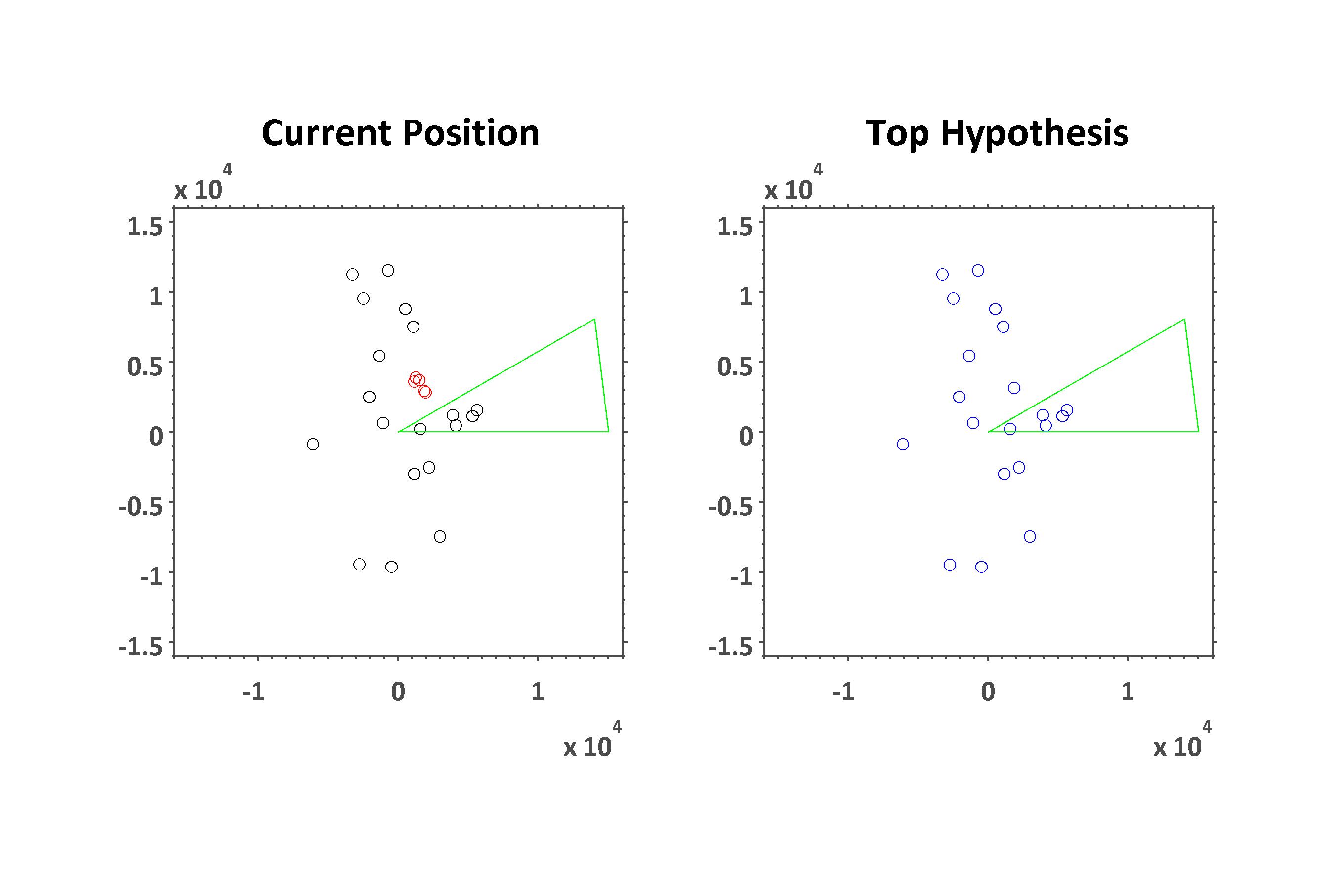}
\label{t2}
}

\subfigure[Partially in the field of view.]{
\includegraphics[scale=.07]{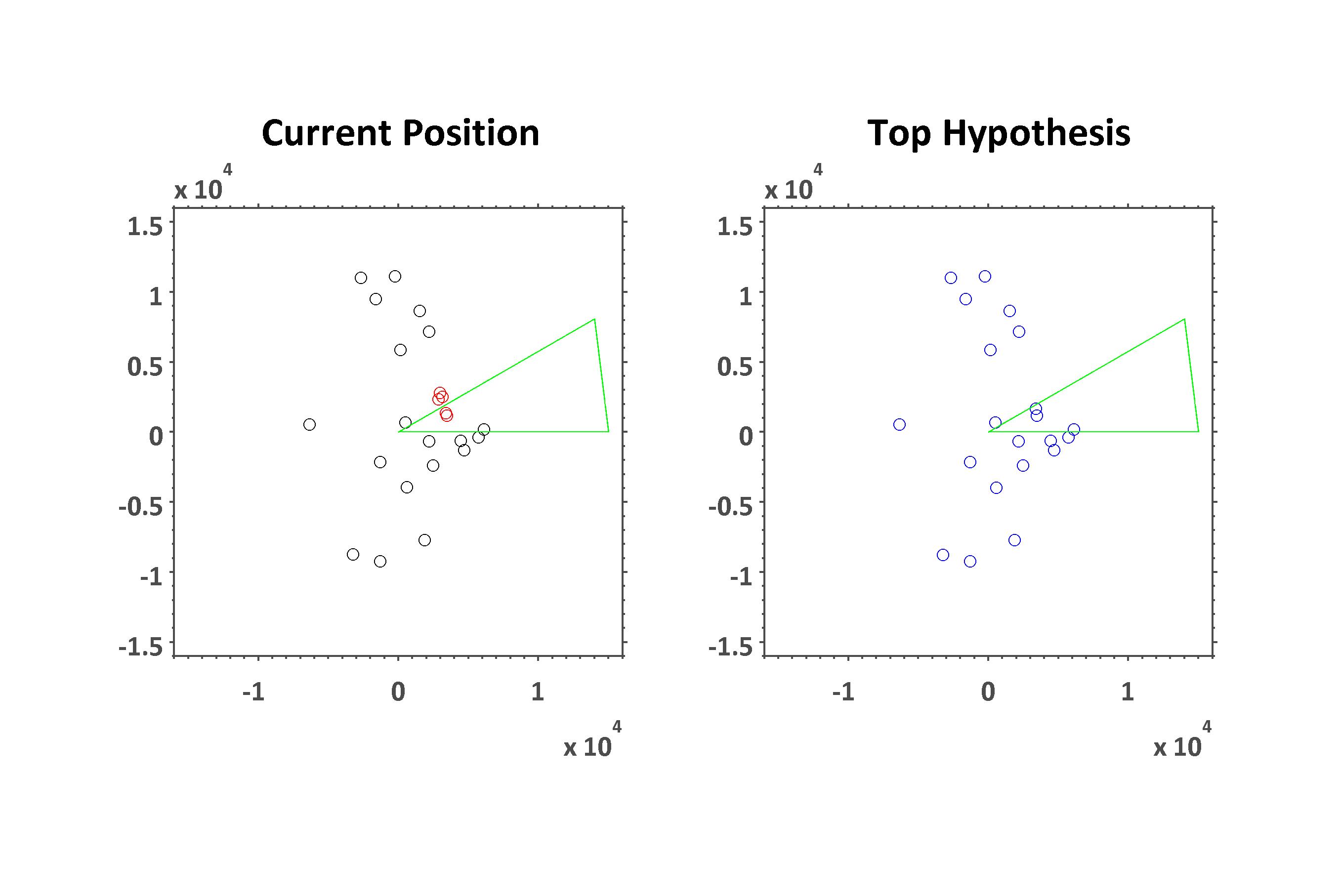}
\label{t4}
}
\end{figure}

\begin{figure}[H]
\centering
\subfigure[Completely in the field of view.]{
\includegraphics[scale=.07]{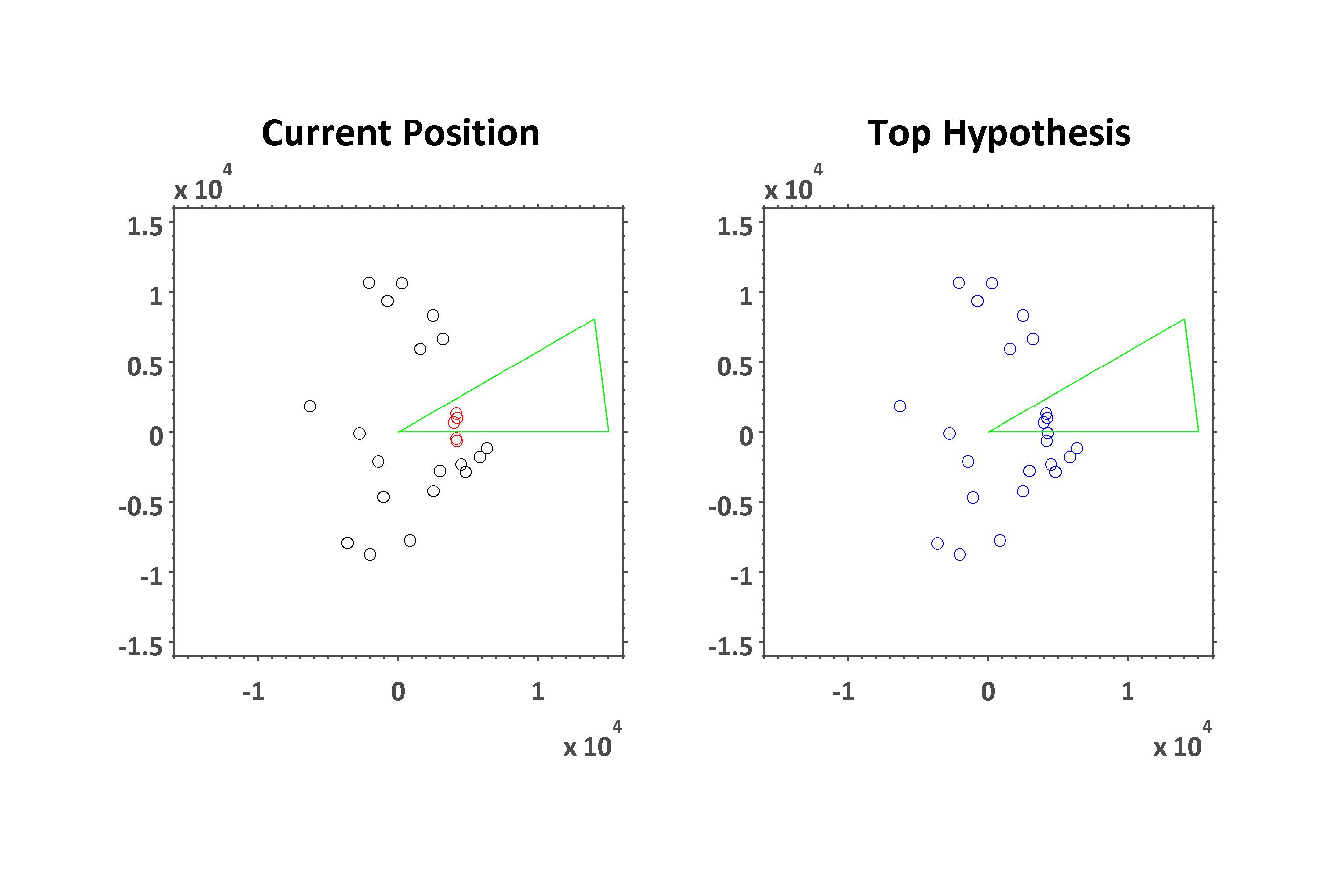}
\label{t6}
}

\subfigure[After passing through the field of view.]{
\includegraphics[scale=.07]{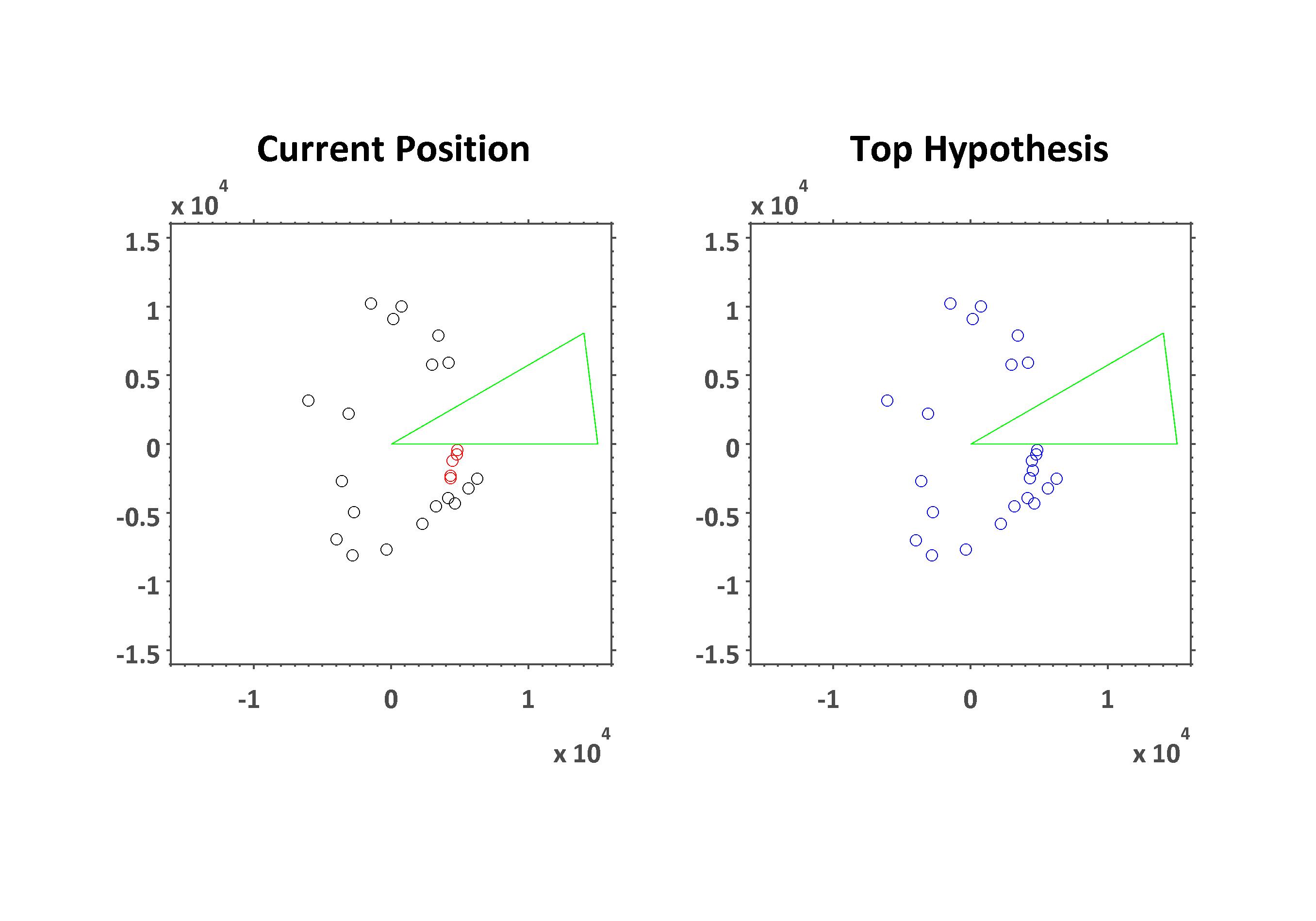}
\label{t8}
}

\caption{Snapshots of the actual states (black and red) and the estimated states from the top hypotheses (green) while passing through the field of view. Axes in tens of thousands of kilometers}
\label{20Spawn}
\end{figure}

We ran a similar simulation to determine if the methodology could capture a birth amongst other non-spawning objects. We did this by simulating a twenty-object SSA example. The initial states for all objects are known with some uncertainty and the same sensor was used. During the simulation one of the twenty objects spawns into multiple objects (shown in red). The goal of this simulation was to accurately track all new object while maintaining hypotheses with the correct number of objects. This simulation is shown in figure \ref{20Spawn}. From figure \ref{t2} one can see that before the spawned object enters the field of view it is still assumed to be intact. As the spawned objects enter the field of view the top hypothesis begins to predict the birth of the spawned objects until all objects are captured. The user defined $\alpha$ and $\beta$ values must be tuned to accurately capture this scenario. In our simulation, heuristics were embedded in the code to adjust the $\alpha$ and $\beta$ values dependent on the ratio between the number of measurements and the number of objects in the field of view. A rigorous approach for adjusting the $\alpha$ and $\beta$ values online is a current topic of our research and will be addressed in future work. 

\begin{figure}[H]
\centering
\includegraphics[scale=.08]{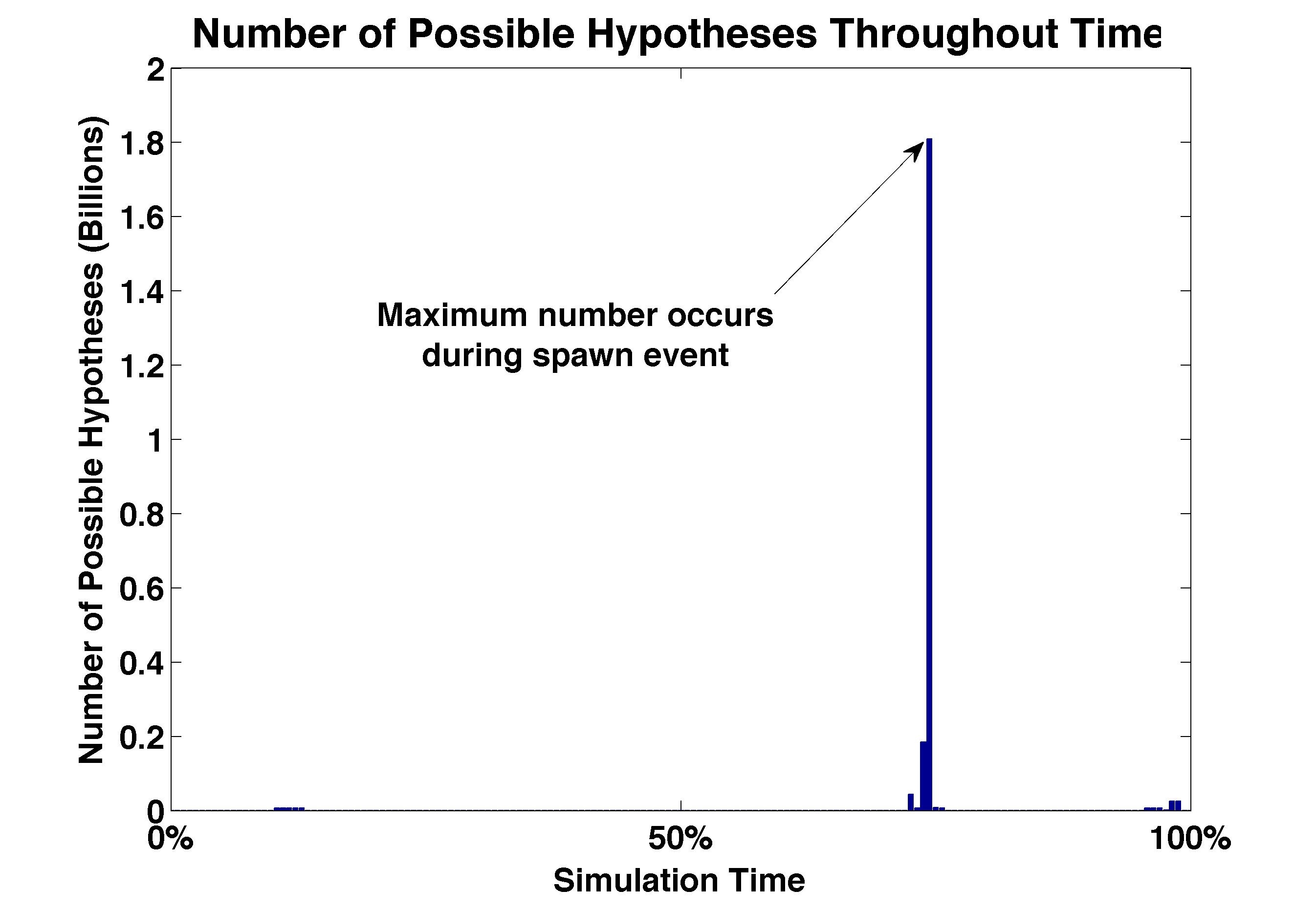}
\caption{Maximum number of possible hypotheses at each step throughout the entire simulation.}
\label{NHs}
\end{figure}

To show the effect of the combination of the spawn event and the ambiguous measurement returns, figure \ref{NHs}
displays the number of possible hypotheses throughout the simulation. 

%\begin{figure}[H]
%\centering
%\includegraphics[scale=.09]{NH}
%\caption{Maximum number of possible hypotheses at each step throughout the entire simulation.}
%\label{NH}
%\end{figure}
Note that the maximum number of possible hypotheses occurs when the spawn event passes through the field of view. This is because the number of possible hypotheses is a function of the number of associable objects and the number of measurement returns. The more ambiguous measurements the greater the number of hypotheses which in this moderate sized case still runs into the billions. %One may notice that the maximum number of hypotheses is still low enough to compute exhaustively. Although this may be true for this particular example as the number of objects and spawn events increases exhaustive computation rapidly becomes intractable. For example, figure \ref{NHs} shows the number of possible hypotheses for a sixty object one spawn event scenario. 
Thus, exhaustively generating these many hypotheses is computationally intractable using conventional methods and emphasizes the need for a computationally tractable approach like RFISST. 
\section{CONCLUSIONS}
In this paper, we have presented a newly developed randomized approach to the multi-object tracking problem called RFISST. RFISST provides a tractable solution to situations were ambiguous measurement returns cause the number of possible hypotheses to increase and become computationally burdensome. We applied the RFISST technique to SSA tracking problems with a random spawn event. The RFISST technique was able to track the spawn event and predict the correct number of objects even though the possible hypotheses were, at times, too numerous to generate exhaustively. In future work we will scale the problem to larger SSA applications, with multiple spawn events.  We will also develop a rigorous approach for adjusting the $\alpha$ and $\beta$ values online, which was done heuristically in this paper. 
\section{ACKNOWLEDGMENTS}
\label{.7}
This work is funded by AFOSR grant number: FA9550-13-1-0074 under the Dynamic Data Driven Application Systems (DDDAS) program.

\bibliographystyle{IEEEtran}
\bibliography{MAP_refs,FISST_refs,RFISST_Refs}

%%%%%%%%%%%%%%%%%%%%%%%%%%%%%%%%%%%%%%%%%%%%%%%%%%%%%%%%%%%%%%%%%%%%%%%%%%%%%%%%

\end{document}